\documentclass[twocolumn,superscriptaddress,showpacs,preprintnumbers,amsmath,amssymb]{revtex4}

\usepackage{epsfig}
\usepackage{graphicx}
\usepackage{dcolumn}
\usepackage{bm}

\begin{document}

\title{Observation results by the TAMA300 detector \\
on gravitational wave bursts from stellar-core collapses }


\author{Masaki Ando} 
\email[e-mail: ]{ando@granite.phys.s.u-tokyo.ac.jp}
\affiliation{Department of Physics, The University of Tokyo, 
     Bunkyo-ku, Tokyo 113-0033, Japan}

\author{Koji Arai}
\affiliation{National Astronomical Observatory, Mitaka, Tokyo 181-8588, Japan} 
\author{Youichi Aso}
\affiliation{Department of Physics, The University of Tokyo, Bunkyo-ku, Tokyo 113-0033, Japan} 
\author{Peter Beyersdorf}
\affiliation{National Astronomical Observatory, Mitaka, Tokyo 181-8588, Japan} 
\author{Kazuhiro Hayama}
\affiliation{Department of Astronomy, The University of Tokyo, Bunkyo-ku, Tokyo 113-0033, Japan} 
\author{Yukiyoshi Iida}
\affiliation{Department of Physics, The University of Tokyo, Bunkyo-ku, Tokyo 113-0033, Japan} 
\author{Nobuyuki Kanda }
\affiliation{Graduate School of Science, Osaka City University, Sumiyoshi-ku, Osaka 558-8585, Japan} 
\author{Seiji Kawamura}
\affiliation{National Astronomical Observatory, Mitaka, Tokyo 181-8588, Japan} 
\author{Kazuhiro Kondo}
\affiliation{Institute for Cosmic Ray Research, The University of Tokyo, Kashiwa, Chiba 277-8582, Japan} 
\author{Norikatsu Mio }
\affiliation{Department of Advanced Materials Science, The University of Tokyo, Kashiwa, Chiba 277-8561, Japan} 
% ...

\collaboration{the TAMA collaboration}


\date{\today}

\begin{abstract}

We present data-analysis schemes and results of observations 
with the TAMA300 gravitational-wave detector,
targeting burst signals from stellar-core collapse events.
In analyses for burst gravitational waves, the detection and 
fake-reduction schemes are different from well-investigated 
ones for a chirp-wave analysis, because precise waveform templates are 
not available.
We used an excess-power filter for the extraction of gravitational-wave
candidates,
and developed two methods for the reduction of fake events
caused by non-stationary noises of the detector.
These analysis schemes were applied to real data from the TAMA300 
interferometric gravitational wave detector.
As a result, fake events were reduced by a factor of 
about 1000 in the best cases.
The resultant event candidates were interpreted from an
astronomical viewpoint.
We set an upper limit of $2.2\times 10^3$ events/sec
on the burst gravitational-wave event rate in our Galaxy
with a confidence level of 90\%.
This work sets a milestone and prospects
on the search for burst gravitational waves,
by establishing an analysis scheme for the observation data from
an interferometric gravitational wave detector.

\end{abstract}


\maketitle

\section{Introduction}

Direct observations of gravitational waves (GWs) will 
reveal new aspects of the universe \cite{bun-thorne1}.
Since GWs are emitted by the bulk motion of matter,
and are hardly absorbed or scattered, they have a potential to 
carry astrophysical and cosmological information 
different from that by electromagnetic waves.
In order to create a new field of GW astronomy,
several groups around the world are developing and 
operating GW detectors.
Among them, much effort is being made recently
on interferometric detectors:
LIGO \cite{bun-LIGO} in U.S.A., 
VIRGO \cite{bun-VIRGO} and GEO \cite{bun-GEO} in Europe,
and TAMA \cite{bun-TAMA1,bun-TAMA3} in Japan.
These detectors have wide-frequency-band sensitivity between 
about 10\,Hz and a few kHz range, and have an ability
to observe the waveform of a GW, which would contain 
astrophysical information.
In these detectors, both high sensitivity and high stability are 
required because GW signals are expected to be extremely faint 
and rare.
 
There are several kinds of target GW sources in these 
interferometric detectors \cite{bun-thorne2,bun-schutz}, 
and data-analysis schemes are being developed and applied to 
the observation data.
Since GW signals are considered to be faint, 
an efficient data-analysis scheme is required to extract 
the remains of GWs from noisy detector outputs.
The target GWs are classified by the signal waveforms:
chirp waves, continuous waves, burst waves, and so on.
A chirp wave is a sinusoidal waveform with increasing 
frequency and amplitude in time, 
which is radiated from an inspiraling compact 
binary just before its merger.
Since this waveform is well-predicted, 
an effective and sophisticated method of matched 
filtering is used in chirp-wave search;
correlations between the data and a template (the predicted waveform)
are calculated to extract a GW signal from noisy data
\cite{bun-BA,bun-Tagoshi}.
A continuous wave has a sinusoidal waveform with a stable frequency 
for over many years,
which is radiated from a quasi-stationary compact binary or
a rotating neutron star.
A matched filtering method can also be used in the search for continuous 
waves from known sources,
because we can predict their waveforms precisely
by observations with electromagnetic waves 
\cite{bun-Niebauer,bun-LIGOcont}.

A burst wave, which is the target of this article, is also
a promising gravitational-waveform class.
This wave has a short spike-like waveform with a duration 
time of less than 100\,msec, which is emitted from 
stellar-core collapse in a supernova explosion or a merger 
of a binary system.
Unlike the chirp and continuous-wave cases, a matched filtering
method cannot be used in a burst-wave analysis.
This is because a set of precise waveform templates that cover the 
source parameters is not available, 
while typical waveforms are obtained by numerical simulations
\cite{bun-SNwave0,bun-SNwave1,bun-SNwave2}.
Thus, 
several signal-extraction methods, called 'burst filters', have been 
proposed for the detection of these burst gravitational waves: 
an excess power filter \cite{bun-excesspow}, 
a cluster filter in the time-frequency 
plane \cite{bun-TFcluster}, 
a slope (or a linear-fit) filter \cite{bun-slope}, and
a pulse correlation filter \cite{bun-cor}.
Since we have only a little knowledge on the waveforms,
these filters look for unusual 
events in the Gaussian-noise background.

For the detection of GWs, evaluation and reduction of 
fake-event backgrounds are critical problems.
In each analysis scheme described above,
we define an evaluation filter 
(such as a correlation with the template in a matched filtering 
method, and certain statistics describing any unusual behavior 
of the data in burst analyses), 
and record the filter output as a GW event trigger 
if it is above a given threshold.
The event triggers usually contain fake events,
which are caused by statistical and externally induced excesses 
in the detector noise level.
Although an ideal interferometric detector would have 
a stationary Gaussian-noise behavior, 
the detector output is far from stationary in practice,
affected by external disturbances, such as seismic motion, 
acoustic disturbance, changes in the temperature and pressure 
and so on.
As a result, it is likely that real signals are buried in these 
fake events, or are dismissed with a larger detection threshold 
set to reduce fakes.
Thus, rejection of these fake events, or a veto in other words, 
is indispensable for the detection of GWs.
In a chirp-wave analysis case, the output of the matched filter
is less affected by non-stationary noises because
it is only sensitive to a waveform similar to GWs.
In addition, since we know a precise waveform of the target GW signal, 
we can reject the fake events by evaluating how well the 
candidate waveform fits to the template
\cite{bun-BA,bun-Tagoshi,bun-BAveto}.
On the other hand, the affects of fake events are more serious 
and critical in the burst analysis case.
Since burst filters are designed to extract any unusual behavior 
of the detector output,
they are also sensitive to non-stationary noises by their nature.
Moreover, it is not straightforward to distinguish these fakes from a 
real signal, and to reject them,
because we do not know the precise GW waveforms.

There are several schemes to reject these fake events:
coincidences by multiple detectors, 
veto analyses with detector monitor signals, 
rejection by waveform behaviors, and so on.
Among them, the most powerful and reliable way
will be a coincidence analysis with multiple independent detectors
\cite{bun-coin1,bun-coin2,bun-coin3}.
If we detect gravitational-wave candidates with multiple detectors
simultaneously (or within an acceptable time difference),
we can declare the detection of a real signal with high confidence.
In a rough estimation, the fake rate is reduced 
by a power of the number of independent detectors.
On the other hand, fake reduction with a single 
detector is also important, even in a coincidence analysis as
the rejection of fakes with a single detector
would reduce accidental coincidences.
In observation runs, many auxiliary signals are recorded 
together with the GW signal-channel in order to monitor the 
detector status.
Since some of them are sensitive to detector 
instabilities, it is possible to reject non-stationary noises
with them \cite{bun-veto2,bun-dia}. 
In addition, even without precise GW waveforms,
fake events are rejected by investigations of the signal behavior 
with our knowledge or assumptions on the waveforms \cite{ref_ngnr}.

In this article, we present a data-analysis scheme for burst GWs,
and results obtained by applying them to real observation data.
The data used in this work were over 2700 hours of data obtained during 
the sixth, eighth, and ninth data-taking runs 
(DT6, DT8, and DT9, respectively)
of the TAMA300 detector \cite{bun-TAMA1,bun-TAMA3,ref_DT8}.
We adopted an excess power filter as a burst filter, 
which is robust for uncertainties of the GW waveforms
\cite{bun-excesspow,ref_ana}.
In addition, we used two fake-reduction methods.
One was a veto with detector monitor signals.
Another was our new method of rejection based on the waveform 
behavior of the time scale.
Although there have been several previous works on similar veto methods
\cite{bun-veto2,bun-dia,ref_ngnr}, 
they were applied to a limited subset of observation data.
We implemented these veto methods in our burst analysis
code, analyzed real observation data, and evaluated 
their effectiveness.
Such a full-scale analysis is important because the effectiveness 
of the vetoes strongly depends on the quality of the real data.

The obtained event triggers were interpreted from
an astronomical point of view;
the results were used to set upper limits on Galactic events.
We carried out Monte-Carlo simulations of Galactic
events with waveforms obtained by numerical simulations
of stellar-core collapses.
In previous works, upper limits by real observations
have been set with artificial waveform models of  
short spikes, Gaussian waves, or sine-Gaussian waves
\cite{bun-coin1,bun-LIGOburst,bun-IGECburst}.
On the other hand, realistic waveforms by numerical simulations
have been used to evaluate the efficiencies of burst filters
with simulated Gaussian noises
\cite{bun-slope,bun-cor}.
We intended to set upper limits in a realistic way:
using a realistic distribution of the Galactic events, 
targeting at realistic waveforms obtained by numerical simulations,
and analyzing long observation data from the detector.
As a result, we expect to obtain
prospects for both current and future research.

This article is organized as follows.
In Section\,II, we overview our burst analysis:
target waves, analyzed data, and our burst filter.
In Section\,III, we describe veto methods 
with a monitor signal and the signal behavior.
After that, we present analysis results and 
an interpretation of the results, setting an upper limit on the
Galactic events, in Section IV.
At last, we present discussions and a conclusion of our
research in Section V and VI.

\begin{figure}[t]
  \begin{center}
  \epsfig{file=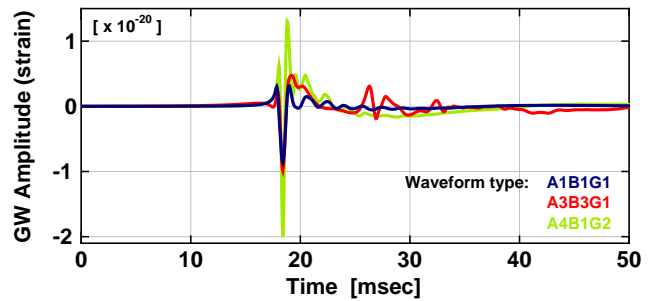, width=8.5cm}
  \caption{Waveform examples from the DFM gravitational waveform catalog.
   The amplitudes are for an event at 10\,kpc distance from the 
   detector.}
  \label{fig-waves}
  \end{center}
\end{figure}

\section{Generation of Event Triggers}

In this Section, we overview the 
target GW signal characteristics, used data, and a burst filter 
used in this work.

\subsection{Target gravitational waves}

The target of the analysis in this work is a burst GW 
from stellar-core collapse (core-collapse supernova explosion).
It is difficult to predict its waveform analytically,
because of the complex time evolution of the mass densities in
the explosion process.
Thus, the explosion process and radiated GWs have been 
investigated by numerical simulations 
\cite{bun-SNwave0,bun-SNwave1,bun-SNwave2}.
Although these simulations were performed with differently 
simplified models, similar waveforms were obtained 
in these simulations.

Among these simulations, Dimmelmeier et al. have presented
rather systematic surveys on GWs from stellar-core collapses
\cite{bun-SNwave0}.
They have obtained  26 waveforms with relativistic simulations
of rotational supernova core collapses, 
with axisymmetric models with different initial conditions 
in a differential rotation, an initial rotation rate,
and an adiabatic index at subnuclear densities.
Although the simulation did not cover all of the initial parameters, 
we used them as reference waveforms in our analysis,
assuming that typical characteristics 
and behavior of the GWs from stellar-core collapses are 
included in this waveform catalog.

\begin{figure}[t]
  \begin{center}
  \epsfig{file=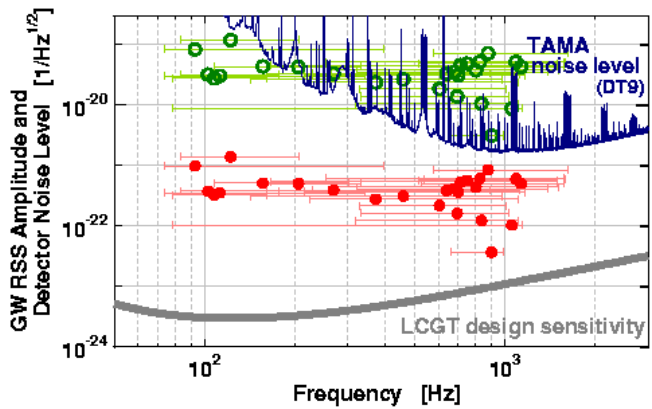, width=8.5cm}
  \caption{RSS amplitude and center frequency calculated from 
      waveforms in the DFM catalog. 
      The amplitudes are for events at the Galactic center
        (closed circles), and for events at 100\,pc distance
        from the detector (open circles).
       Each error bar indicates the frequency range
       within which the spectrum density value is above half
       of its peak value.
       The noise level of TAMA at DT9 and the design sensitivity of 
        LCGT \cite{bun-LCGT} are shown together.}
  \label{fig-waves_spe}
  \end{center}
\end{figure}

We processed the original waveforms of the catalog 
(we call it DFM catalog in this article) with a 30\,Hz
second-order high-pass digital filter, and resampled them to
20\,kHz in order to be compatible with the data from the detector
(described in the next part).
Figure\,\ref{fig-waves} shows examples from the waveform catalog.
While these waveforms have different behaviors,
they have common characteristics: about a 1\,msec-short spike,
and a total duration of less than 100\,msec.
According to the DFM catalog, the averaged amplitude 
of GWs radiated by supernovae at the Galactic center 
(8.5\,kpc distance from the detector)
is $ \langle h_{\rm peak}\rangle = 1.5 \times 10^{-20} $
in a peak strain amplitude,
or $ \langle h_{\rm rss} \rangle 
=4 \times 10^{-22}\ [{\rm Hz^{-1/2}}]$
in root-sum-square (RSS) amplitude.
Here, a RSS amplitude
is defined by
\begin{equation}
   h_{\rm rss} = 
   \left[{ \int_{-\infty}^\infty \left|{ h (t)}\right|^2 \ dt}\right]^{1/2},
\end{equation}
where $h(t)$ is the strain amplitude of the GW \cite{bun-LIGOburst}.
The central frequencies of the waves range from 90\,Hz to 1.2\,kHz
(Fig.\,\ref{fig-waves_spe}),
which is around the observation band of interferometric detectors.
Also, it is estimated from the DFM catalog that
a total energy radiated as GWs in one event is
$ \langle E_{\rm tot}\rangle = 8 \times 10^{-8}\, [M_\odot c^2]$, 
in average.
Here, $ M_\odot$ is the mass of the Sun.

\subsection{Data from a gravitational wave detector TAMA300}

\begin{table}[b]
  \begin{center}
    \caption{Summary of long data-taking runs by TAMA300.
    The noise level and total observation data amount are described.
     The last column (D.C.) represents the duty cycle throughout
     the data-taking run.}
    \label{tb-DTR}
    \begin{tabular}{ c c c c c}
      \hline
          & Term & Noise level  & Total data & D.C. \\
          &      & [${\rm Hz^{-1/2}}$] & [hours] & \\
      \hline
        DT6 & Aug. - Sept, 2001 &  $ 5\times 10^{-21}$ & 1038 & 87\% \\
        DT8 & Feb. - April, 2003 & $ 3\times 10^{-21}$  & 1157 & 81\% \\
        DT9 & Nov., 2003 - Jan., 2004 & $ 2\times 10^{-21}$ & 558 & 54\%\\
      \hline
    \end{tabular}
  \end{center}
\end{table}

We applied our analysis method to observation data obtained 
by TAMA300 \cite{bun-TAMA1,bun-TAMA3}; 
TAMA300 is a Japanese laser-interferometric 
gravitational wave detector, located at the Mitaka campus 
of the National Astronomical Observatory of Japan (NAOJ) 
in Tokyo (${\rm 35^{\circ} 40' N,\ 139^{\circ} 32' E}$).
TAMA300 has an optical configuration of 
a Michelson interferometer with 300\,m-length 
Fabry-Perot arm cavities and with power recycling to enhance
the laser power in the interferometer.
During the operation, the mirrors of the detector are 
shaken by a 625\,Hz sinusoidal signal,
which enables us to calibrate the detector sensitivity continuously
with a relative error of less than 1\% \cite{bun-cal}.
The main output signal of the detector, which would contain
GW signals, is recorded with a 20\,kHz, 16 bit
data-acquisition system \cite{bun-daq}.
Besides the main output signal, 
over 150 monitor signals are also recorded during the observation:
signals for the laser power in and from the interferometer, 
detector control-loop signals, seismic and acoustic 
monitor signals, signals for temperature and pressure monitor, 
and so on \cite{ref_DT8}.
These monitor signals are used for diagnosing the detector
condition, and for veto analyses
(Section \ref{sec-veto}).
The recorded data are stored in DLT tapes on site, and are 
sent to data-analysis computers at the collaborating 
institutes by Giga-bit optical network connections.

\begin{figure}[t]
  \begin{center}
  \psfig{file=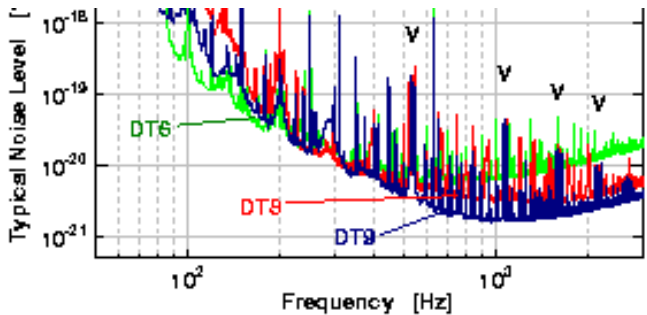, width=8.5cm}
  \caption{Typical noise level during data-taking runs with
   the TAMA300 detector.
   The noise level has been improved run by run.
   The spectrum contains several line peaks: harmonics of 50\,Hz AC power line,
    violin mode peaks of the 
    suspension wire of the mirror (described as 'V'), 
    and a calibration peak at 625\,Hz (described as 'C').}
  \label{fig-sens}
  \end{center}
\end{figure}

In TAMA, nine observation runs have been carried out so far
since the first observation run in 1999,
and over 2700 hours of data have been collected.
In this work, we used the data in the sixth, eighth, and 
ninth data-taking runs (DT6, DT8, and DT9, respectively, 
Table\,\ref{tb-DTR}).
We obtained over 1000 hours of data in each of DT6 and DT8,
operating the detector stably and with a good duty cycle.
The duty cycle in DT9 was not very high,
because most of the day time was spent for the adjustment and 
measurement of the detector during the first half term.
On the other hand, we obtained data with
uniform quality with a high duty cycle in the second half,
which included that during the quiet time of new-year holidays.
The duty cycle was 96\% (207 hours' observation in 216 hours) 
in this quiet term of DT9.
We used only this stable term as the DT9 data in the following analyses.

Typical noise spectra in these observation runs are shown 
in Fig.\,\ref{fig-sens}.
The noise level has gradually been improved by 
detector investigations between these runs.
The detector has a floor-level sensitivity of around from 300\,Hz to 2\,kHz
frequency range.
The floor level is $ 2\times 10^{-21}\, {[\rm Hz^{-1/2}]}$
in DT9 at around 1\,kHz.
The spectrum contains several line peaks: harmonics of a 50\,Hz AC line,
violin mode peaks (around 520\,Hz and integer multiples) of the 
suspension wire of the mirror, 
and a calibration peak.
Since these lines could affect the analysis results,
they were removed in the data analyses.

\subsection{Extraction of signals by an excess-power filter}

We developed and implemented a burst-wave analysis code
based on an excess-power burst filter. 
Among several filters proposed so far,
an excess-power filter  \cite{bun-excesspow}
and a TF-cluster filter \cite{bun-TFcluster}
look for an increase of power in the data of a detector,
while a slope filter \cite{bun-slope}
and a pulse correlation filter \cite{bun-cor}
monitor correlations between the data and assumed waveforms.
Roughly speaking, a higher detection efficiency is attained
with assumptions on the waveform.
On the contrary, the efficiency is drastically degraded 
if there are any errors in the assumed waveforms.
%
An excess-power filter is robust because it uses 
only a little information on 
the target waveforms: 
the signal duration time and the frequency band.
The evaluation parameter is
the total noise power in a given time-frequency region.
In spite of its robustness, 
it is nearly as efficient as matched filtering
for signals with short duration and a limited frequency band
\cite{bun-excesspow}.

\begin{figure}[t]
  \begin{center}
  \psfig{file=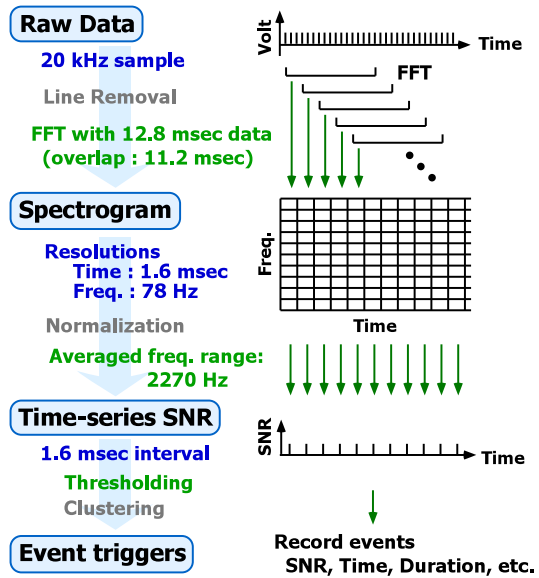, width=7cm}
  \caption{Data-processing chart of our excess-power burst filter.
     We first calculate a spectrogram from the detector output data.
     Next, we obtain the time-series SNR by averaging the frequency
     components.
     Then, we extract event triggers by a given threshold.}
  \label{fig-excess}
  \end{center}
\end{figure}

Our filter generates event triggers in the following steps 
(Fig.\,\ref{fig-excess}, details are described in Appendix\,\ref{sec-append}):
(i) A spectrogram (time-domain change in noise power spectrum)
is calculated from the output data of the detector;
the power spectrum is calculated with a $\Delta t =12.8$\,[msec] 
data chunk using a fast Fourier transform (FFT), 
which is repeated with 1.6\,msec time-delays.
We used a Hanning window in each FFT process to obtain
smooth spectra and time-series results.
Here, each spectrum has a frequency resolution of 
$1/\Delta t = 78$\,[Hz].
Since the original data contains many line peaks 
(AC line peaks in every 50\,Hz, etc.),
this low-resolution spectrum is contaminated by these peaks.
Thus, these peaks are removed from the original time-series 
data before calculating the spectrogram. 
(ii) In each spectrum, power in pre-selected frequency bands are 
averaged so as to obtain a time-series of averaged power, $P_n$.
Each spectrum is normalized (whitened) by the typical noise 
spectrum within 30\,min before a calculation of the average 
in the frequency components.
As a result, $ P_n$ represents the signal-to-noise ratio (SNR): 
the ratio of the averaged signal power 
to the typical noise power in pre-selected time-frequency region.
In this work, we selected a fixed band of $ \Delta f=2270$\,[Hz]
from 230\,Hz to 2.5\,kHz.
(iii) Event triggers are extracted if the averaged power is 
larger than a given threshold, $ P_n \geq P_{\rm th}$.
If unusual signals in the detector output
are sufficiently large, 
they will be observed in the filter output.
Continuous excesses above the threshold are clustered 
to be a single event.
Each event trigger is recorded with its parameters:
the peak signal power $P_{\rm ev}$, the time of the event $t_{\rm ev}$,
the duration time above the threshold, and so on.

The parameters of the filter, length of the time chunk 
($\Delta t$) for each FFT, and analysis frequency band ($\Delta f$)
were selected to be effective for the reference burst GW signals.
According to the DFM catalog,
the signals have short spike-like waveforms,
i.e. a short duration and a wide 
frequency band.
Although the selected parameters 
($ \Delta t = 12.8$\,[msec], $ \Delta f = 2270$\,[Hz])
were not fully optimized for the waveforms,
the analysis results were not changed very much
with a different parameter set.
Moreover, 
we could keep the robustness of the excess-power filter
with this rough tuning of the time-frequency bands.

\subsection{Signal-injection simulations}
\label{sec-soft-inj}

\begin{figure}[t]
  \begin{center}
  \psfig{file=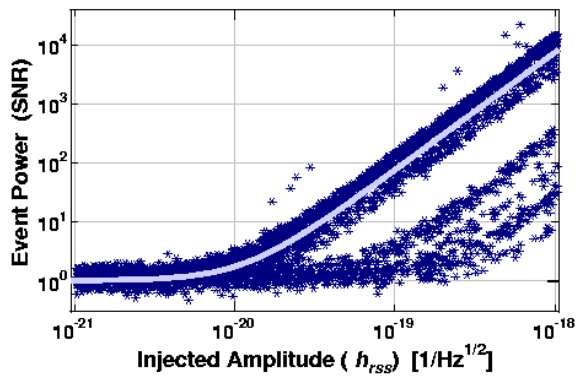, width=7.5cm}
  \caption{Relation of injected signal amplitude and
       SNR obtained by the injection test for the DT9 data.
      The asterisk points and the curve represent the signal-injection 
      results and the fitting result, respectively.
      The points at the lower right side of the plot are caused by 
      waveforms in which the signal power are concentrated 
      at a low-frequency band; 
      the sensitivity of TAMA is worse in these frequency bands
      (see Fig.\,\ref{fig-waves_spe}).}
  \label{fig-inj1}
  \end{center}
\end{figure}

The output of the filter ($ P_{\rm ev}$) is a dimensionless signal-to-noise 
ratio (SNR).
SNR is calibrated to a physical value,
such as the GW amplitude, by results of
signal-injection simulations (called 'software injection tests')
with real data from the TAMA300 detector.
In the simulation, target reference waveforms are 
superimposed to the detector data with proper calibration
(estimated from the transfer functions of the detector, 
the whitening filter, the anti-aliasing 
filter before the data-acquisition system).
The signals are injected and analyzed one by one
with equal time separations in the data so as to evaluate the data uniformly.
The amplitudes and waveforms were selected randomly
from $ 10^{-22} \leq h_{\rm rss} \leq 10^{-18}\,[{\rm Hz^{-1/2}}]$
and from 26 waveforms in the DFM catalog, respectively.
This data were analyzed by the same code as that for the 
raw data analysis.

The results of the signal-injection test are shown in 
Fig.\,\ref{fig-inj1};
the recorded power SNRs of the events ($ P_{\rm ev}$)
are plotted as a function of the root-sum-square amplitudes 
of the injected signal ($h_{\rm rss}$).
The injection results of each data-taking run were fitted by
\begin{equation}
  P_{\rm ev} = 1 + (C_{{\rm DT}x} \times h_{\rm rss})^2, 
  \label{eq-SNR}
\end{equation}
where $C_{{\rm DT}x}$ represent the averaged efficiency 
coefficients ($x$: the data-taking number).
From the injection results, we obtained the coefficient values: 
$ C_{\rm DT6}= 2.2 \times 10^{19}$,
$ C_{\rm DT8}= 3.3 \times 10^{19}$, 
and $ C_{\rm DT9}= 8.7 \times 10^{19}$.

An inverse of this coefficient corresponds to the GW amplitude with
which the signal power is the same as the noise power by our filter.
Thus, it is interpreted as the noise-equivalent amplitude
of the GW signal.
The noise-equivalent GW amplitude was 
$ h_{\rm rss, noise}=1.1 \times 10^{-20}\, [{\rm Hz^{-1/2}}]$
for DT9 with our excess-power filter.
From the estimation that the averaged GW amplitude was 
$ \langle h_{\rm rss}\rangle= 4\times 10^{-22}\,[{\rm Hz^{-1/2}}]$ 
for a 8.5\,kpc event, 
TAMA has the ability to detect events
within around 300\,pc away from Earth with this 
noise-equivalent amplitude.

\section{Reduction of fake events}
\label{sec-veto}

In this Section, we describe veto methods to
reject fake events caused by detector instabilities.
We have used two veto methods:
a veto method using auxiliary signals for the detector monitor,
and a veto method by the 
waveform behavior: the time-scale of the signal.

\subsection{Veto with auxiliary signals for the detector monitor}
\label{sec-veto1}

Here, we describe a veto method using 
auxiliary signals recorded together with the main 
output of the detector:
a correlation between the monitor signal and the main 
output, the confirmation not to reject real GW signals, 
and a false dismissal rate estimation.


\subsubsection{Veto with the intensity monitor signal}

\begin{figure}[t]
  \begin{center}
  \psfig{file=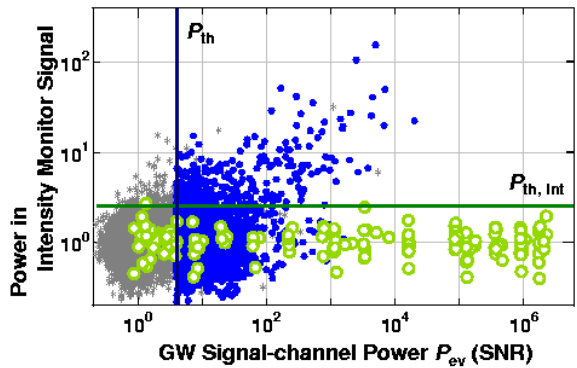, width=7.5cm}
  \caption{Correlation between the powers in the 
    main output channel and the intensity monitor channel.
    The closed circles are event candidates selected by the 
    threshold $ P_{\rm th}=4$;
    the gray asterisks are powers at the time 100\,msec-shifted from the
    event candidates, 
    and the open circles are results of hardware-injection tests.}
  \label{fig-int1}
  \end{center}
\end{figure}

We investigated some of the monitor signals, and found 
strong correlations between the short spikes
in the main output and the monitor signal for the laser 
intensity in a power-recycling cavity of the interferometer. 
Figure\,\ref{fig-int1} shows the correlation between 
the main output and the intensity monitor signal.
The intensity-monitor data were processed by 
the same excess-power filter to detect short-spike instabilities.
The filter parameters were the same as that for 
the main signal analysis, except for the frequency range.
In the analysis of the intensity-monitor data, the frequency
range was $ \Delta f_{\rm int} =1170$\,Hz from the DC frequency,
which was determined from the spectrum shape of the burst spikes in 
the intensity data.
The closed circles in Fig.\,\ref{fig-int1} represent
the power (SNR) of the events in the GW data-channel 
and in the intensity monitor.
In this figure, the event triggers were extracted with a 
threshold of $ P_{\rm th}=4$ for the DT8 data.
For a comparison, the powers at the time 100\,msec-shifted from the
triggers are also plotted in this figure (gray asterisks)
\footnote{We have confirmed that correlations in the filter
outputs were sufficiently small with time shift of 100\,msec.}.
This figure shows that there are strong correlations between 
these two signal powers for some of the event triggers,
and only weak correlations outside of the event triggers.

Thus, vetoes of fake events with the intensity monitor signal 
are expected to be effective;
when the outputs of two excess power filters (one for the GW signal-channel, 
and another for the intensity monitor) exceed the respective
thresholds simultaneously, 
the triggers are labeled as fakes, and 
are removed from the event candidate list.

\subsubsection{Estimation of a false-dismissal rate}

To use the veto with the intensity monitor signal, 
we should confirm that the intensity instabilities were not 
caused by huge GW signals;
otherwise, we may reject real GW signals by this veto.
During DT8, we investigated the response of the detector
by injecting simulated waveforms to it.
In this test, which is called a 'hardware injection' test,
we shook the interferometer mirrors with
a short-spike waveform and a typical burst-waveform 
obtained by numerical simulations 
\cite{bun-SNwave1} with various amplitudes.
The results of the hardware-injection test are plotted 
as open circles in Fig.\,\ref{fig-int1}.
There were 147 injections, and 117 events were above
the event-selection threshold of $ P_{\rm th}=4$.
We observed no excess power in the intensity monitor
with an intensity veto threshold of $ P_{\rm th, int}=2.2$
(described below), 
while the injected signals appeared with sufficiently large powers 
in the filter output for the GW signal-channel.
The number of accidental excesses with this threshold 
is expected to be 1.2 (1\% of injected events).
Thus, the result that no excess was found above the intensity threshold
is well-consistent with the expected accidental background.
From these results, we ensured that the intensity instabilities were not 
caused by huge GWs, and that it is safe to use the intensity 
monitor signal for a veto analysis.

\begin{figure}[t]
  \begin{center}
  \psfig{file=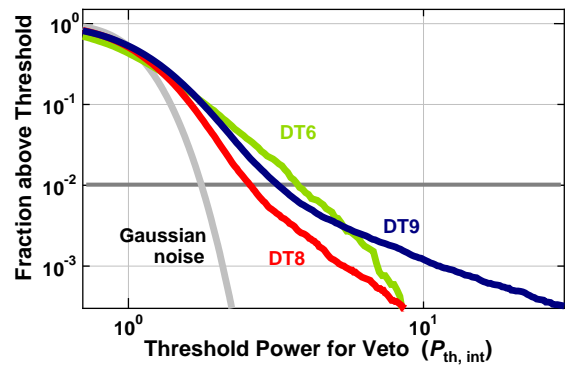, width=7.5cm}
  \caption{Estimation of an accidental coincidence.
    The intensity-veto threshold was decided so that the
    accidental coincidence probability 
    (or the false-dismissal rate in this case) would be 1\%.
    As a result, the thresholds are
     $ P_{\rm th, int}= 3.9$, 2.2, and 3.0 for DT6, DT8, and DT9,
     respectively.}
  \label{fig-int2}
  \end{center}
\end{figure}

The threshold for the intensity excess power is selected
so as to reduce fake events effectively with small 
probability to reject real GW signals, i.e. with a 
small false-dismissal rate.
The false-dismissal rate is equal to the accidental coincidence rate 
between the intensity excess and the excess in GW signal-channel,
which was estimated from the distribution of the 
power in the intensity monitor signal.
We selected a threshold so that 
the false-dismissal rate would be 1\%,
which resulted in thresholds of
$ P_{\rm th, int}= 3.9$, 2.2, and 3.0 for 
DT6, DT8, and DT9, respectively  (Fig.\,\ref{fig-int2}).
The difference in the thresholds for these data-taking runs were
caused by the difference in the detector stability in each run
and the improvement of the intensity monitor instrument 
between DT8 and DT9.

\subsection{Veto by a signal behavior test}
\label{sec-veto2}

Here, we describe the second veto method,
a veto method by the time-scale of the signal.
Statistics for the signal evaluation,
and an estimation of the false-dismissal rate 
are described.

\subsubsection{Signal behavior test}

As described above, fake events are rejected by careful 
selection and an investigation of the monitor signal-channels.
However, it is hard to see any clear correlations
for all of the fake events in practice,
because there are various origins of the fakes, which are
difficult to be identified.
Thus, a test of the signal behavior at the main output of 
the detector will be helpful to reduce fake events.

The effectiveness of the veto with a signal behavior test depends
on how well we know, or how many assumptions we set, 
on the signal behavior.
In the burst-wave analysis case, the waveforms by numerical 
simulations suggest that GWs from stellar-core collapse 
have a short duration, typically less than 100\,msec.
We know that some of the detector instabilities last longer
than a few seconds from experience.
Thus, some of the fakes caused by these slow instabilities 
are rejected by evaluating the time scale of the event 
triggers \cite{ref_ngnr}.

\subsubsection{Evaluation statistics for the time-scale veto}

\begin{figure}[t]
  \begin{center}
  \psfig{file=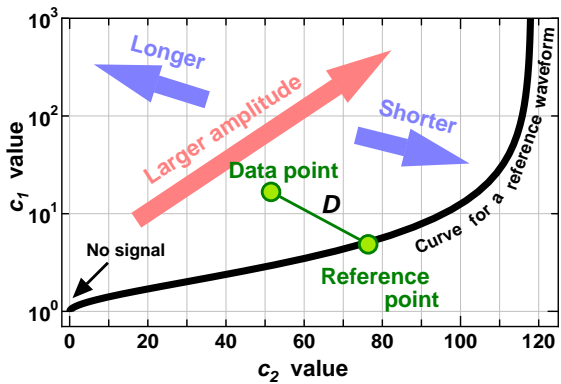, width=7.5cm}
  \caption{Evaluation of a data point on a $c_2$-$c_1$ correlation plot.
         The curve shows the expected locus for a reference 
         waveform, which was drawn by sweeping the signal amplitude.
         The distance between a data point and a reference point on
         the reference curve represents the similarity of the
         signal amplitudes and their time scales.}
  \label{fig-model}
  \end{center}
\end{figure}

In this work, we selected to evaluate the time 
scale of event triggers with statistics around the event
(Appendix\,\ref{sec-appendB} for details).
The excess-power filter outputs $ P_n$,
a time series of the power in a pre-selected frequency band.
We calculate two statistics from 
the $ \pm \Delta T/2$ time-series data around the event-candidate 
time $ t_{\rm ev}$:
\begin{equation}
  c_1 = Q_1\ \ \ {\rm and}\ \ \ 
  c_2={1 \over 2}\left({ { Q_2 \over  Q_1^2}-1 }\right),
\end{equation}
where $ Q_k=\sum P_n^k /N $ ($k$: integer) is
the $k$-th-order moment of the power 
\cite{bun-kur, bun-NRC}. 
Here, $ N $ is the number of power-data points in
the evaluation time (within $ t_{\rm ev} \pm \Delta T/2$).
The statistics $ c_1 $, which is related to an averaged power, 
has information about the stability of the noise level
on a given time scale.
On the other hand, $ c_2$ is related to the second-order
moment of the noise power.
Since it is normalized by the averaged power ($ Q_1$),
the $ c_2 $ value becomes constant
if the signal power 
is much larger than the background noise level.
The constant number 
is determined by the time scale of the event: 
large in a short-burst case, 
and small in a case of a slow change in the noise power.

\begin{figure}[t]
  \begin{center}
  \psfig{file=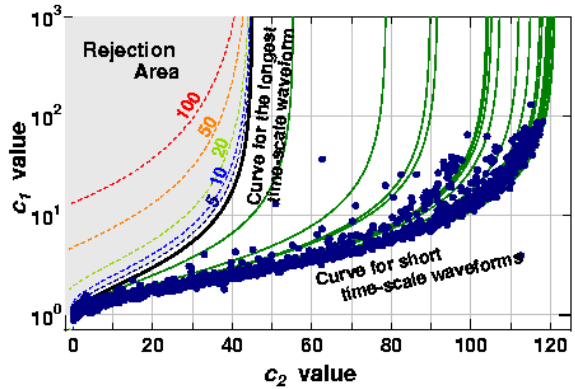, width=7.5cm}
  \caption{Correlation plot of $c_1$ and $c_2$ used for 
         the rejection of fake events, calculated with
         a time window of $ \Delta T=0.82 $\,[sec].
         The solid curves show the expected loci for the reference 
         waveforms from the DFM catalog.
         Among them, the curve for the waveform with the longest 
         time-scale is shown as a bold curve.
         The dotted curves represent the contours for the distance 
         $D_{\rm min}$ to the longest time-scale reference curve.
         When event triggers are in the gray area, which is 
         the $ D_{\rm min}\geq 5$ area, they are rejected as fakes.
      The closed circles represent the results of the signal-injection test 
       using the data of DT9.
        With a fake-event-selection threshold of $ D_{\rm th}=5$,
        the false dismissal rate was estimated to be 0.3\%.}
  \label{fig-table}
  \end{center}
\end{figure}

We use an evaluation parameter ($D_{\rm min}$), which
represents the similarity to the GW signal,
estimated from the $c_1$ and $ c_2$ statistics.
When we plot the data point in the $ c_2 $-$ c_1$ plane 
(Fig.\,\ref{fig-model}),
they will be in the left region for long-duration event 
cases and in the right region for short burst-like event cases,
and will be in the upper region for large power events.
Thus, two independent information on each event,
the power and the time scale, are shown by the position 
of each data point in this plot.
Here, the distance ($D$) between the data point and a 
reference point (expected position of the real GW signal, 
estimated from the signal waveform), 
represents the similarity of their signal behaviors.
The evaluation parameter ($D_{\rm min}$) is the smallest
distance for all of the reference waveform and amplitude combinations.

In a practical application of this method to real data,
we set a loose selection criteria for fake events
 (Fig.\,\ref{fig-table}).
We have 26 reference GW waveforms from the DFM catalog, 
which have different time scales.
Instead of comparing the time scale of an event trigger
with that of each reference waveform, we compare it only with 
the longest time scale of the reference GW waveforms.
In other words, the evaluation parameter of an event
($D_{\rm min}$) is set to be zero if its time scale
is shorter than that of the longest reference waveform.

\subsubsection{Selection of parameters for the time-scale veto}

Fake events with different behaviors from that of the real ones
are rejected when the minimum distance ($D_{\rm min}$) is
larger than a given threshold ($ D_{\rm th}$).
Here, two parameters should be set in this veto analysis:
an evaluation time-window ($\Delta T$) and
a fake-selection threshold ($ D_{\rm th}$).
The time window is selected to be $ \Delta T = 0.82 $\,sec,
so that the veto would be effective.
Since this time window length is between the typical time scales of 
the fakes (larger than a few seconds) and 
the real GW signals (less than 100\,msec),
we can expect a clear distinction between them.
On the other hand, the threshold for the rejection ($ D_{\rm th}$) 
is selected so that the false-dismissal rate of real 
GW signals would be acceptable.

The false-dismissal rate was directly evaluated from the 
results of a signal-injection test with the real data 
from the TAMA300 detector
(described in Sec.\,\ref{sec-soft-inj}).
This simulation is important because the real data from a 
detector do not have an ideal Gaussian noise distribution.
The closed circles in Fig.\,\ref{fig-table} show the results of the 
signal-injection test to the DT9 data.
The injection results are distributed well-around the 
theoretical predictions shown as solid curves.
With a fake-event-selection threshold of $ D_{\rm th}=5$,
the false-dismissal rate was estimated to be 0.3\%;
6 injection results were rejected out of 2006 injections
\footnote{The estimated false-dismissal rate depends on the 
distribution of the waveforms and amplitudes of the injected signals.
In the Galactic signal-injection test described in 
Section\,\ref{sec-galinj}, the false-dismissal rate was only 0.08\%.}.
Although this value is larger than that estimated by a 
statistical analysis, it is sufficiently small for a veto
analysis.
The false dismissal rates were also investigated
for the DT6 and DT8 data with a similar signal-injection test, 
and found to be 0.6\% and 2.9\%, respectively.
The differences come from the original behavior of the data
in these observation terms.

\begin{figure}[t]
  \begin{center}
  \psfig{file=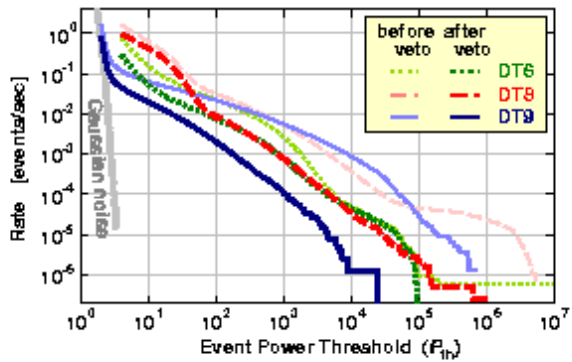, width=7.5cm}
  \caption{Event-trigger rate results with TAMA data.
      The horizontal and vertical axes represent the threshold 
      ($ P_{\rm th}$) and the trigger rate in a unit of events/sec, 
      respectively. 
      The analysis result with simulated Gaussian noise is also 
      plotted together with the DT6, DT8, and DT9 results.}
  \label{fig-rate1}
  \end{center}
\end{figure}

\section{Data-processing results with the TAMA300 data}
\label{sec-res}

The analysis method described above was applied to 
real data from TAMA300.
In this section, we consider the results of the TAMA
data analysis with the vetoes, and the interpretation of the 
results from an astronomical point of view.
The data were mainly processed by a PC-cluster computer
placed at the University of Tokyo.
This machine is comprised of 10 nodes, and has 20 CPUs
(Athlon MP 2000+ by AMD Inc.).
The analysis time for the excess power filter was about 
30-times faster than the real time;
it took about 1/30 sec to process 1-sec data.

In the data processing, the first 9-min and the last 1-min data 
of the each continuous observation span were
not used because they sometimes contained loud noises
caused by detector instabilities, or excited violin-mode 
fluctuations.
In addition, the duration time of rejected fake events 
is considered as a dead time of the detector, 
and subtracted from the total observation times.
The dead times by the fake rejections were
1.3\%, 1.7\%, and 0.4\% of the observation time
for DT6, DT8, and DT9, respectively.
The effective observation times ($T_{\rm obs}$) 
are shown in Table\,\ref{tb-sum}.

\subsection{Event-trigger rates}

\begin{figure}[t]
  \begin{center}
  \psfig{file=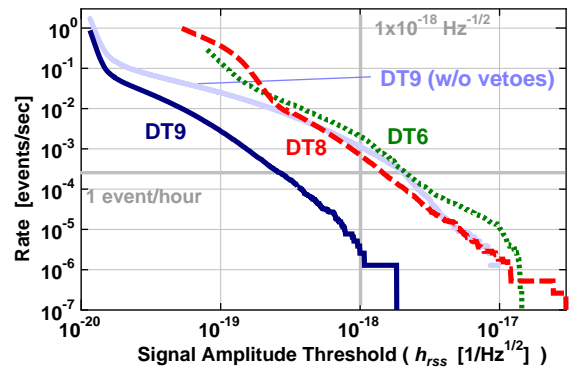, width=7.5cm}
  \caption{GW amplitude and corresponding trigger rate;
     the event rate (vertical axis) with larger amplitude than a given 
     $h_{\rm rss}$ (horizontal axis) is plotted.}
  \label{fig-rate2}
  \end{center}
\end{figure}

Figure\,\ref{fig-rate1} shows the event-trigger rates obtained by 
the TAMA data analyses;
the trigger rate (in a unit of events/sec) is plotted as a 
function of the event-extraction threshold ($ P_{\rm th}$). 
The analysis result with simulated Gaussian noise is also 
plotted in Fig.\,\ref{fig-rate1}, together with the DT6, DT8, and
DT9 results.
Assuming that the real GW signals are rare and faint,
we can regard most of the triggers as being fakes.
From this figure, one can see that the trigger rates were 
reduced in these data-taking runs with vetoes.
For a given GW power threshold ($P_{\rm th}$), 
the trigger rates were reduced by 1/10-1/1000.
The power threshold could be reduced 
(for better GW-detection efficiency) 
by a factor of 10-100 for a given trigger rate.
Figure\,\ref{fig-rate2} shows the event-trigger rates
plotted as a function of $h_{\rm rss}$ amplitude,
which was obtained from Eq.\,(\ref{eq-SNR}).
The detector was gradually improved during the intervals 
of these data-taking runs.
The event-trigger rates were reduced from DT6 to DT9
by about a few orders for a given GW amplitude,
and by about an order for given trigger rates
(Table\,\ref{tb-sum}).

\begin{table}[b]
  \begin{center}
    \caption{Summary of data analysis results.
       The noise-equivalent GW RSS-amplitudes ($ h_{\rm rss, noise}$), 
       the dead times by the vetoes ($ T_{\rm rej}$),
       the total effective observation times ($ T_{\rm obs}$),
       the trigger rates for $ h_{\rm rss} \geq 
        1\times 10^{-18}\, [{\rm Hz^{-1/2}}]$, and 
        the GW RSS-amplitudes above which the trigger rates are 
        one event per hour are described.}
    \label{tb-sum}
    \begin{tabular}{ c c c c c c}
      \hline
         & $ h_{\rm rss, noise} $ & $ T_{\rm rej}$  & $ T_{\rm obs}$ 
         & Rate & 1-${\rm hour^{-1}}$ amp. \\
         &  [$ {\rm Hz^{-1/2}} $] & [hours] & [hours] &  [${\rm sec^{-1}}$] 
         & [$ {\rm Hz^{-1/2}} $] \\
      \hline
        DT6\ \ & $ 4.5\times 10^{-20}$ &  11.8 &  937.8  
            & $ 2.1 \times 10^{-3} $ & $ 2.1\times 10^{-18}$ \\
        DT8\ \ & $ 3.0\times 10^{-20}$ & 18.0 & 1064.2  
            & $ 7.0 \times 10^{-4} $ & $ 1.4\times 10^{-18}$ \\
        DT9\ \ & $ 1.1\times 10^{-20}$ &  0.8 & 194.6 
            & $ 2.5 \times 10^{-6} $ & $ 2.5\times 10^{-19}$\\
      \hline
    \end{tabular}
  \end{center}
\end{table}


\begin{figure}[t]
  \begin{center}
  \psfig{file=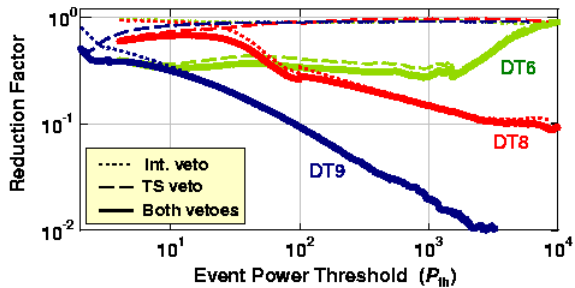, width=7.5cm}
  \caption{Reduction factor by two veto methods for DT6, DT8, and DT9.
    The bold curves are the reduction ratio with two vetoes.
    The dotted and dashed curves are only with an intensity veto
    and a time-scale veto, respectively. }
  \label{fig-rej_rate}
  \end{center}
\end{figure}

Figure\,\ref{fig-rej_rate} shows the reduction factor
of event triggers with two veto methods;
the ratio of the number of event triggers 
after and before the vetoes are plotted
as a function of the SNR threshold.
In these runs, both of the two vetoes contributed 
to the reduction of the rates.
They worked in complementary ways.
The intensity veto is effective to short-duration 
fakes and large SNR fakes.
On the other hand, the time-scale veto is effective 
for the long-time instability of the detector output and small 
SNR fakes.

In DT9, many event triggers were rejected as fakes by
the intensity veto.
This is because the pre-amplifier and whitening filter 
for the data acquisition of the intensity signal were 
improved in this run.
Since the reduction factor is better for large
SNR fakes, it is expected that the reduction ratio will be
further improved with a higher detection efficiency of the
intensity instabilities.
On the other hand, only a small fraction of fakes
were rejected by the time-scale selection in DT9
because the detector operation was sufficiently stable.
The detector was operated very stably thanks to a quiet
seismic environment during the holiday 
weeks in the second half of DT9.
In addition, the drift of the typical noise level was 
small at that time.
In DT6, the time-scale veto was much more effective 
than the intensity veto.
This is because DT6 data contained noisy data originating in 
an instability of the laser source and seismic disturbances 
during the daytime.

The rates are still much larger than that
with simulated Gaussian noises, even with
the vetoes and the improvements in the detector.
In addition, the trigger rate is still much higher than the
expected rate of supernova explosions.
The expected GW event rate is one event in a few tens of 
years, i.e. about $10^{-9}$\,events/sec in our Galaxy.
(Here, note that TAMA has an ability to detect only 
events within 300\,pc away from Earth at best.)
These results suggest that most of the observed trigger-events were 
fake events caused by an instability of the
detector, even with vetoes.

\subsection{Simulations of Galactic events}
\label{sec-galinj}

Since the event-trigger rate is still much larger than that 
expected from ideal Gaussian noise or observed supernova rates,
we cannot claim the detection of GW signals from the 
data-analysis results.
Thus, we set upper limits for stellar-core collapse events 
in our Galaxy.
We carried out Monte-Carlo simulations with a source-distribution 
model of our Galaxy, and with waveforms from
the DFM catalog.
The simulated data were analyzed in the same way as the 
detector data, and compared with the observation results.

In the simulation, we adopted a source-distribution model 
based on the observed luminous star distribution in our Galaxy,
assuming that the event distribution of the stellar-core collapses
was identical to it.
There have been studies on the star-distribution model 
based on sky-survey observations \cite{ref-galmod1,ref-galmod2}.
In our simulation, we used a simple axisymmetric distribution 
model (an exponential disk model) described in a 
cylindrical coordinates,
\begin{equation}
   \rho (R, \theta, z) 
    \propto \exp \left({-{R \over R_0} - {|z| \over h_0}}\right),
    \label{eq-gal-dist}
\end{equation}
where $\rho$, $ R_0 = 3.5$\,kpc, and $ h_0=320 $\,pc are 
the density of the events, and the characteristic
radius and height of the density of 
the Galactic disk, respectively.
As well as the non-axisymmetric components, such as 
spiral arms, the thick disk and halo structures were
neglected in our simulations because their number of stars was
only about 3\% of that of the disk component \cite{ref-galmod2}.
We adopted $ R_\odot = 8.5$\,kpc and $ h_\odot = 20$\,pc
for the position of the Sun in our simulation.

We used 200\,hours stable data in the second half of DT9 
for the Galactic signal-injection test.
This test was performed according to the following steps:
(i) Set the GPS times at which simulated events are injected; 
   these times are uniformly separated between the start and 
   end times of the observation run.
  Decide the position of each event randomly according to 
  the Galactic-event distribution described by 
  Eq.\,(\ref{eq-gal-dist}).
  Select the waveform of each event randomly from the 
  DFM catalog.
 (ii) Calculate the distance and sky direction 
     seen from the detector for each event, from
    the position of the event in the Galaxy and
    the injection time information.
 (iii) The expected GW amplitude is calculated from the distance
   to the event source and the detector antenna pattern 
    for the sky position of the event
   \footnote{Note that we only consider the orientation dependence 
      of the detector;
      the orientation dependence of the source is not considered.
      In other word, we assumed that the GWs were radiated
      with maximum amplitude toward the detector.}.
    We assumed non-polarized GWs; 
    the GW power is equally distributed
    to the two polarizations.
  (iv) Inject each event waveform to the TAMA300 data
    with estimated amplitude, and analyze the data with a 
    similar code as that for the raw-data analysis.
  (v) Extract the events at the injected time.

\subsection{Results of Galactic-event simulations}

\begin{figure}[t]
  \begin{center}
  \psfig{file=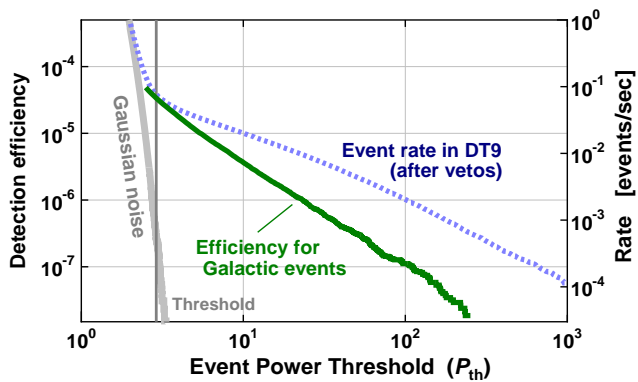, width=8.5cm}
  \caption{Results of a Galactic-event simulation.
      The solid curve shows the detection efficiency for the 
       Galactic events as a function of the threshold.
      The trigger rate obtained by the DT9 observation 
      is also plotted as a dotted curve 
       for a comparison (right vertical axis).}
  \label{fig-gal-res1}
  \end{center}
\end{figure}

Figure\,\ref{fig-gal-res1} shows the results of a 
Galactic-event simulation. 
The fraction of the detectable Galactic events
(the detection efficiency, $ \epsilon_{\rm gal}$, left axis) 
is plotted as a function of the SNR threshold ($P_{\rm th}$).
The event-trigger rate in DT9 is also plotted for a comparison
(right axis).
With an event-selection threshold of $ P_{\rm th}=3.0$
(which corresponds to averaged amplitude of 
$ h_{\rm rss, th}= 1.6 \times 10^{-20}\,[{\rm Hz^{-1/2}}]$
for DT9), the detection efficiency was estimated to be
$ \epsilon_{\rm gal} = 3.1 \times 10^{-5} $
for the Galactic events.
The threshold was selected so that the expected contribution 
of the Gaussian noise would be sufficiently small (less than 1\% of the
triggers above the threshold).
The upper limit for the event rate 
determined from the TAMA raw-data was 
$ R_{\rm DT9, UL} = 6.8 \times 10^{-2} $\,events/sec
with a confidence level of 90\%.
From these results, we obtained the upper limit for the 
Galactic event rate to be 
$ R_{\rm gal, UL} = R_{\rm DT9, UL}/\epsilon_{\rm gal}
 = 2.2 \times 10^{3}  $\,[events/sec]
with a 90\% confidence level
\footnote{
Slightly better upper limit was obtained by a limiting
SNR range to be analyzed.
When we concentrated on the events with SNRs between 3.9 
and 4.65, the event upper limits became
$ R_{\rm gal, UL} = 1.5 \times 10^{3}  $\,[events/sec]
and $ \dot{E}_{\rm GW, UL} = 2.6 \times 10^{-4}\, [M_\odot c^2$/sec].}.
This value is considerably larger than
the theoretical expectation of about $10^{-9}$\,events/sec.

Besides the upper limit for the rate of a stellar-core 
collapse in our Galaxy, an upper limit was set for
the GW energy rate.
The total energy radiated as GW, $ E_{\rm tot}$, was 
estimated for each event from its waveform.
The upper limit for the energy rate radiated as 
GW was estimated by the product of the event-rate upper limit,
$ R_{\rm gal, UL} $, and the averaged GW energy
of the events, $ \langle E_{\rm tot}\rangle$.
As a result, we obtained 
$ \dot{E}_{\rm GW, UL} = 3.8 \times 10^{-4}\, [M_\odot c^2$/sec].
Again, this value is considerably large;
the rate of the total energy radiated as GWs would be about
$ M_{\rm Gal}/(2\times 10^7) \,[M_\odot c^2$/years],
where $ M_{\rm Gal} $ is the total mass of our Galaxy, 
which we assume to be $2\times 10^{11}\,M_\odot$.

There are uncertainties in setting the upper limits
by several origins.
Here, we consider the effect of the detector calibration error, 
statistical error in the Monte-Carlo simulation 
for the Galactic events, and the error in the Galactic model.
The calibration error in the conversion of the detector output
to the GW strain amplitude was estimated to be less than 1\%.
This calibration error causes an amplitude error in the 
signal-injection test.
From Eq.\,(\ref{eq-SNR}) and the results of the Galactic 
signal-injection test, the uncertainty in the upper limit is 
estimated to be 2.9\% with a detector calibration error of 1\%.
On the other hand, the statistical error in the Monte-Carlo simulation
is determined by the number of the simulated events above the
threshold.
We generated $ 3.8\times 10^8 $ Galactic events, and
detected $1.2\times 10^4$ events above the threshold of 
$ P_{\rm th}=3.0$.
Assuming that the number of detected events follows a Poisson
distribution, the event-rate uncertainty is 0.9\%.
At last, the error in the Galactic model would affect the results.
The parameter $ R_0$ in the Galactic model 
has an error of 9.4\% \cite{ref-galmod2}.
This error corresponds to a 5.8\% uncertainty in the detection efficiency
for the Galactic events ($\epsilon_{\rm gal}$)
with a threshold of $ P_{\rm th}=3.0$, which was estimated
by additional simulations.
In total, the uncertainty in our upper-limit results is 6.5\% at most.
The detection efficiency will be reduced
(the upper limit will be increased) by 1.2\% by including
a thick disk and halo components.

\section{Discussions}

\subsection{Comparison with previous studies}

We interpreted the observation results
from the viewpoint of the Galactic event rate in the previous section.
In this part, we interpret the results in an similar way as in the
previous studies for a comparison;
we set an upper limit on the rate of GW events incident
on the detector as a function of the GW amplitude 
\cite{bun-LIGOburst,bun-IGECburst}.

\begin{figure}[t]
  \begin{center}
  \psfig{file=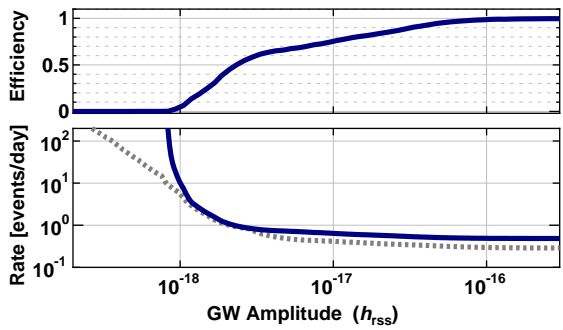, width=7.5cm}
  \caption{Detection efficiency and an upper limit 
   (90\% confidence level) 
   for the waveforms of the DFM catalog.
    The detection efficiency with a given threshold of 
    $ P_{\rm th}=10^4$ is shown in the upper plot,
    and the upper limit on the event rate is shown in 
    the lower plot (solid curve).
    The gray dotted curve is the envelope of the upper 
    limits with various detection thresholds.}
  \label{fig-upper1}
  \end{center}
\end{figure}

We used 200 hours of data in the DT9 stable term,
and set an upper limit on the rate of the events received by the detector,
following the procedure to set the upper limits
by the LIGO group \cite{bun-LIGOburst}.
At first, the detection threshold was fixed, and the upper limit
was set at the number of the events above the threshold with
a given confidence level.
Then, the detection efficiency for a given GW amplitude was
estimated by a software signal-injection test.
Here, the events were distributed randomly on the celestial sphere
in order to include the directivity of the detector.
We investigated the Gaussian (with a time scale of 
$\tau=0.5$ and 1\,[msec]) 
and sine-Gaussian waveforms (with Q-value of 9 and central
frequencies of 554, 850, and 1304\,Hz) 
for a comparison with the previous studies, as well as the waveforms
from the DFM catalog.
Finally, we estimated the upper limit on the event rate
from the upper limit on the number of events, $ N_{\rm UL}$, 
the detection efficiencies, $\epsilon$, and the observation time, 
$ T_{\rm obs}$, by
$ R_{\rm UL}=N_{\rm UL}/(\epsilon \cdot T_{\rm obs})$.

\begin{figure}[t]
  \begin{center}
  \psfig{file=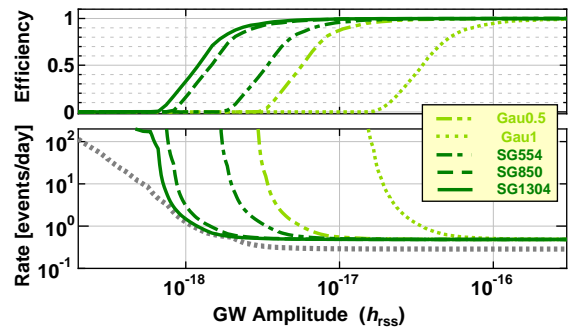, width=7.5cm}
  \caption{Detection efficiencies and upper limits 
  (90\% confidence level) 
  for the Gaussian (time scale of 0.5 and 1\,msec) and sine-Gaussian
  (central frequency of 554, 850, and 1304\,Hz) waveforms.
      The gray dotted curve is the envelope of the upper 
    limits for the 1304\,Hz sine-Gaussian waveform 
    with various detection thresholds.}
  \label{fig-upper2}
  \end{center}
\end{figure}

We set the threshold to be $ P_{\rm th}=10^4$,
which resulted in one trigger above the threshold.
Assuming Poisson statistics, we obtained the corresponding upper 
limit of 3.89 events with a confidence level of 90\%.
The detection efficiencies and the upper limit results are shown 
in Fig.\,\ref{fig-upper1} 
(waveforms from the DFM catalog) and Fig.\,\ref{fig-upper2} 
(Gaussian and sine-Gaussian waveforms).

The upper limit for sufficiently large events
(ex. $h_{\rm rss} > 1\times 10^{-16}\,[{\rm Hz^{-1/2}}]$)
was 0.49 events/day with a confidence level of 90\%.
This upper limit is comparable to the LIGO-S1 result 
of 1.6 event/day \cite{bun-LIGOburst}
and the Glasgow-MPQ coincidence result of 0.89 events/day
\cite{bun-coin1}.
On the other hand, the resonant detector network has set 
an upper limit of $\sim 4\times 10^{-3}$ events/day 
\cite{bun-IGECburst}.
These differences in the upper limits come mostly from the 
different observation times.
As for the sensitivity for smaller amplitude signals, 
the GW amplitude for 50\% detection efficiencies
(averaged over source directions)
was around $ 1\times 10^{-18}\,{\rm Hz^{-1/2}} $ 
for short and high-frequency waveforms in our case.
The upper limit curve is almost comparable with the LIGO-S1 
results for high-frequency signals, and larger
for lower frequency ($<$ 800\,Hz) ones \cite{bun-LIGOburst}.

\subsection{Outlook for the detection of burst GWs}

The large event rate and upper limit results show that the detector 
output is still dominated by fake events, 
even after these vetoes.
Thus, further research efforts are necessary to 
detect burst gravitational waves.
In this part, we discuss the outlook to better vetoes, 
coincidence analyses with other observatories, 
and better performance of the detector and
data-processing scheme.

In Section\,\ref{sec-veto1}, we presented a veto analysis 
method with only one monitor signal,
an intensity monitor.
Similar methods can be used with the other monitor signals
along with careful investigations of their correlations
with the main output of the detector.
However, we have found no other monitor signal with a clear 
correlation so far: laser power at the signal-port 
(the dark port), monitors for the seismic 
fluctuations, an acoustic monitor signal.
Thus, it is necessary to investigate more deeply 
the monitor signals, and to introduce better monitor 
signals that are sensitive to the detector instabilities.

There are other event-selection criteria than
the time-scale selection method
presented in Section\,\ref{sec-veto2}.
For example, the time scale of an event can be simply evaluated by
the duration time above the event-selection threshold.
In this case, we should consider that the veto results will be 
strongly dependent on the event amplitude
\footnote{Veto with band-limited RMS amplitude used in
\cite{bun-LIGOburst} corresponds to a veto only with $c_1$.
In these methods, we cannot avoid huge GW events from being rejected
by the veto.}.
We will be able to reduce fake events even further by knowing 
the common characteristics of the target events,
and setting them as event-selection criteria.
For this, more systematic and precise simulations of stellar-core
collapses and investigation on the waveform will be helpful.

Coincidence analyses with the other detectors for
GWs, electromagnetic waves, and neutrinos will 
improve the result significantly,
though we have focused on the reduction of fakes with 
a single detector in this article.
The observation runs by TAMA300 (DT8 and DT9) were carried out 
at the same term as the LIGO second and third scientific
(observation) runs (called S2 and S3), and coincidence analyses 
are underway  \cite{bun-coinLIGO}.
We note that our work discussed in this article 
is also a part of the LIGO-TAMA coincidence analysis;
the list of the event triggers obtained in our work will 
be used in the coincidence analysis.

In addition to a reduction of fakes, the improvements of 
the detector both in the floor noise level and in the reduction of 
non-stationary noises are also important.
The performance of the TAMA300 detector has gradually been improved 
from DT6 to DT9 concerning both the noise level and the stability,
and the detector still has room for improvement.
In addition, burst filters with higher efficiencies are
under development in the TAMA group and other groups.
Since we can only observe events within about the 300\,pc range
with the current sensitivity of TAMA, the detection efficiency
for the Galactic events is very small 
($ \epsilon_{\rm gal} = 8.9 \times 10^{-5}$ with a threshold
for a noise-equivalent GW amplitude).
The sensitivity should be improved by about two orders 
so as to cover our Galaxy, and to realize a sufficiently large
detection efficiency.
This sensitivity will be realized by the next-generation detectors, 
such as LCGT (Fig.\,\ref{fig-waves_spe}) \cite{bun-LCGT}  
and advanced LIGO \cite{bun-AdLIGO}.

\section{Conclusion}

We presented data-analysis schemes and results
of observation data by TAMA300, 
targeting at burst signals from stellar-core collapses.
Since precise waveforms are not available for burst gravitational 
waves, the detection schemes (the construction of a detection filter 
and the rejection of fake events) are different
from those for chirp wave analyses.
We investigated two methods for the reduction of non-stationary 
noises, and applied them to real data from the TAMA300 
interferometric gravitational wave detector.
As a result, these veto methods, a veto with a detector monitor signal
and a veto by time-scale selection, worked efficiently
in a complimentary way.
The former and the latter were effective for short-spike noises and 
for slow instabilities of the detector, respectively.
The fake-event rate was reduced by a factor of about 1000 
in the best case.

The obtained event-trigger rate was interpreted from the viewpoint
of the burst gravitational-wave events in our Galaxy.
From the observation and analysis results, we set an upper limit
for the Galactic event rate to be $2.2\times 10^3$ events/sec
(confidence level 90\%),
based on a Galactic disk model \cite{ref-galmod2} 
and waveforms obtained by numerical simulations
of stellar-core collapses \cite{bun-SNwave0}.
In addition, we determined the upper limit for the rate of the energy
radiated as gravitational-wave bursts to be 
$ 3.8 \times 10^{-4}\,M_\odot c^2$/sec (confidence level 90\%).
These large upper limits show that the detector output was
still dominated by fake events, even after the selection of events,
and gives us prospects on both current and future research:
the necessity for further improvement of the analysis schemes, 
coincidence analyses with multiple detectors,
better predictions on the waveforms,
and future detectors, such as LCGT and advanced LIGO, 
to cover the whole of our Galaxy.
This work has set, we believe, a milestone 
for these research activities.

\begin{acknowledgments}
This research is supported in part by a Grant-in-Aid for Scientific 
Research on Priority Areas (415) of the Ministry of Education, 
Culture, Sports, Science and Technology.
\end{acknowledgments}

\appendix

\section{Detailed description on the excess-power filter}
\label{sec-append}



\subsection{An excess-power filter}

In this part, we detail the excess-power filter.
We assume that the output of a detector is comprised 
of an ideal stationary Gaussian noise, $ n(t) $, 
and a signal, $ s(t)$ 
(non-Gaussian component caused by gravitational 
waves or instability of the detector):
$ v(t)=n(t)+s(t) $.
The power spectrum is calculated for every $\Delta t$ data chunk
with given time-delays ($\delta t$), 
using a fast Fourier transform (FFT).
As a result, we obtain a spectrogram (a time-frequency plane) 
of the noise (or signal) power with $ \delta t $ time 
separation and $1/\Delta t$ frequency resolution.
Here, the Fourier component is also described by 
the sum of the noise and the signal:
$ \tilde{v}_{mn}=\tilde{n}_{mn}+\tilde{s}_{mn} $,
where $ m$ and $ n $ represent the indices for the 
frequency and time, respectively.
Then, the power in each time-frequency component is described 
by $ |\tilde{v}_{mn} |^2$.

\begin{figure}[t]
  \begin{center}
  \psfig{file=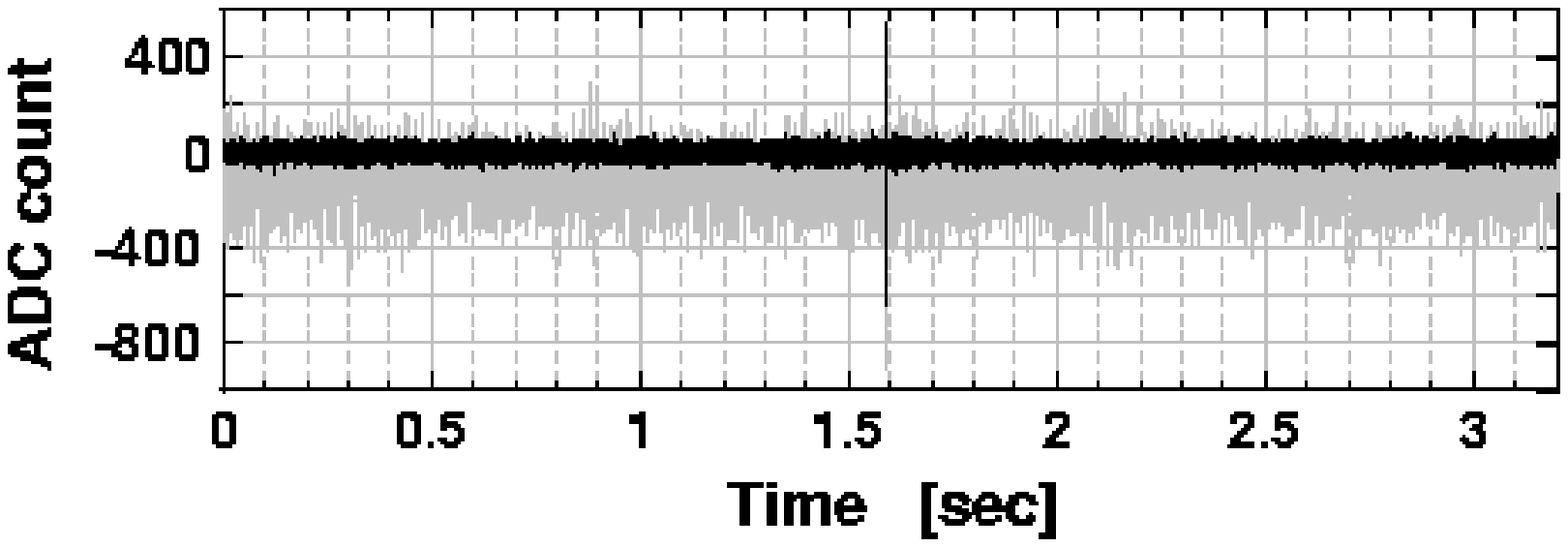, width=7.5cm}\\
  \vspace*{2mm}
  \psfig{file=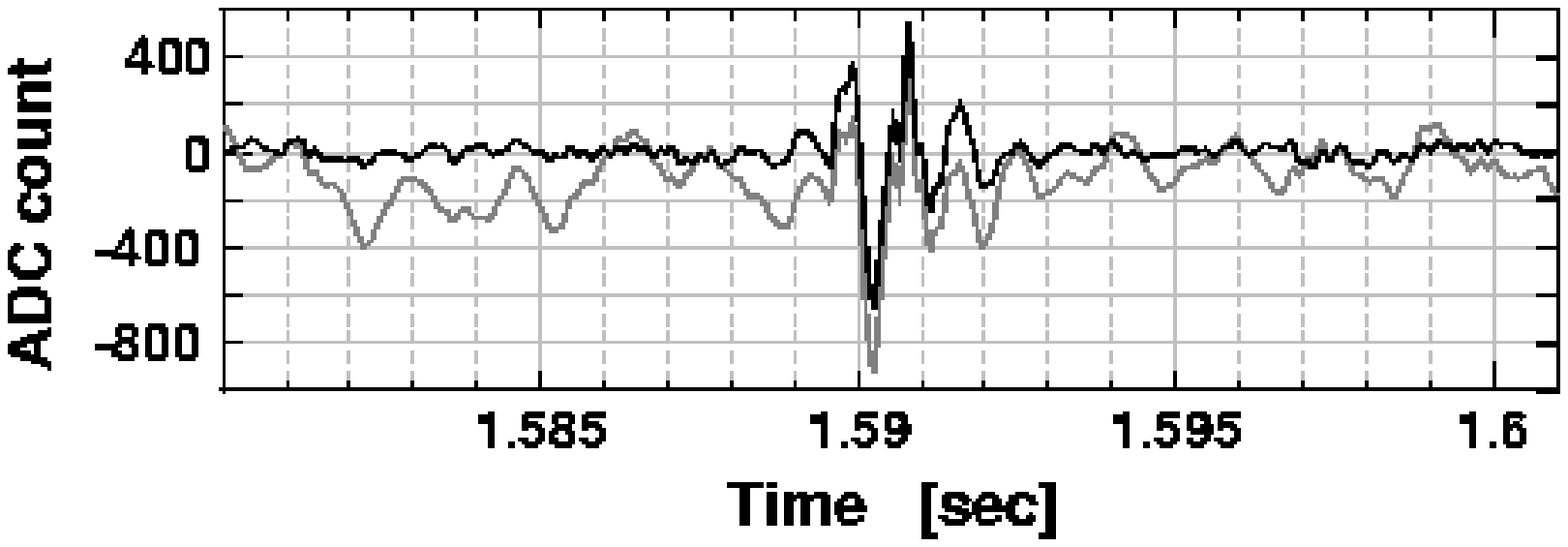, width=7.5cm}
  \caption{Example of the line removal results for the DT9 data.
      Time-series data before (plotted in gray) and after 
      (plotted in black) the line removal are shown.
      The lower plot is a zoom up of the spike in the upper plot.}
  \label{fig-rm1}
  \end{center}
\end{figure}

In order to make our filter equally effective for all of the 
analysis frequencies, we normalize the power spectrum
by an averaged noise level,
$ \overline{N_{m}} =  \sum_{n=n_0}^{n_0+N_{\rm av}-1} 
 |\tilde{n}_{mn}|^2 / N_{\rm av}$,
where $N_{\rm av}$ is the number of time components used for the average.
Then, the normalized power is written as
\begin{equation}
   P_{mn}=  N_{mn} +2 C_{mn} + S_{mn} ,
   \label{eq-P_mn}
\end{equation}
where $ N_{mn}=|\tilde{n}_{mn}|^2/\ \overline{N_{m}} $, 
$S_{mn}= |\tilde{s}_{mn}|^2/\ \overline{N_{m}} $, and 
$ C_{mn}= \Re\{\tilde{s}_{mn} \cdot \tilde{n}_{mn}^* \}/\ 
\overline{N_{m}} $,
meaning a normalized noise power, 
a normalized signal power, and 
a normalized correlation between the signal and noise components,
respectively.
Since $ \tilde{n}_{mn}$ has a Gaussian noise distribution, 
$N_{mn}$ has a $ \chi^2$ distribution of two degrees of freedom
(an exponential distribution):
$  P(N_{mn})= \exp ({-N_{mn}})$.

Then, the output of the excess-power filter,
the averaged power for a given time component, is written as
\begin{equation}
   P_n= {1\over M}\sum_{m} P_{mn},
\end{equation}
where $ M $ is the number of frequency components used in the average;
only the values of the power in pre-selected frequency components are used
to calculate the averaged power, $P_n$.
From Eq.\,(\ref{eq-P_mn}), $ P_n $ is written as
\begin{equation}
   P_n=  N_n  +2 C_n + S_n,
   \label{eq-P_mn_sum}
\end{equation}
where $  N_n \equiv \sum_{m} N_{mn}/M $, 
$   C_n \equiv \sum_{m} C_{mn}/M $, and
$   S_n \equiv \sum_{m} S_{mn}/M $.
This represents the time evolution of the power in detector output
in given frequency bands.

\subsection{Data conditioning}

The data from the detector is not an ideal Gaussian 
noise, in practice; the noise spectrum is not white,
the noise level changes in time,
and many line peaks are included in the output. 
Thus, data conditioning before processing the excess power 
filter is indispensable.

\begin{figure}[t]
  \begin{center}
  \psfig{file=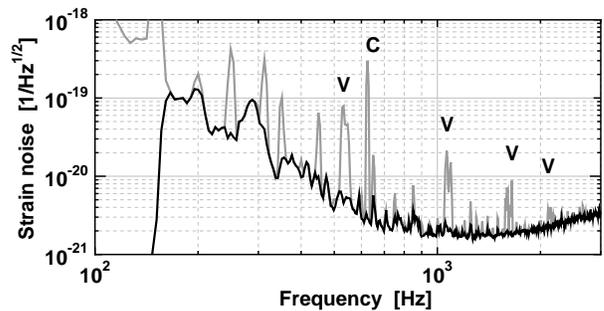, width=8cm}
  \caption{Results of line removal for the DT9 data.
      The noise spectrum before (plotted in gray) and after 
      (plotted in black) the line removal are shown.}
  \label{fig-rm2}
  \end{center}
\end{figure}

The spectrum contains several line peaks: 
harmonics of 50\,Hz AC line, violin mode peaks 
(around 520\,Hz and integer multiples) of the 
suspension wire of the mirror, 
and a calibration peak.
These line peaks are removed in the following processes.
At first, we obtain a Fourier spectrum from 72\,sec of data by FFT.
We then set the line-frequency components to be zero.
In addition, the lower frequency components below 160\,Hz are also rejected.
At last, we obtain a time-series data by calculating the inverse 
FFT of the spectrum.
With this process, the line peaks are clearly removed from the spectrum
(Fig.\,\ref{fig-rm1}, \ref{fig-rm2}).
Moreover, since only a small number of frequency components are rejected,
the burst waveforms are not vary much affected by the 
line removal process.

The frequency and time dependences of the noise spectrum are
compensated by the normalization with the averaged noise power 
spectrum, $ \overline{N_{m}} $.
We calculated $ \overline{N_{m}} $ by averaging 
the power spectra for 30\,min before the data analyzed 
by the excess-power filter.
In order to avoid the large spikes from disturbing the averaged spectrum,
we rejected noisy 0.7\% spectra from those used in each average.
We found that we could obtain stable averaged spectra, and that each
spectrum was normalized well with this method.

\section{Time-scale evaluation and veto}
\label{sec-appendB}


\subsection{Evaluation in $ \Delta${\it T} time chunk}

In this part, we describe the details of the veto method with
time-scale evaluation of the event triggers.
In our veto methods, each event is evaluated by the statistics
in a $ t_{\rm ev} \pm \Delta T/2 $ data chunk 
($t_{\rm ev} $: the time of the event).
Here, $ N$ data points of the excess-power-filter output are contained
in the time window $ \Delta T$, 
i.e. $ \Delta T = N \delta t$.
From the output of the excess-power filter, $ P_n $, 
we define the evaluation parameters
$ c_1$ and $ c_2$ as
\begin{equation}
  c_1 \equiv Q_1,\ \ \ 
  c_2 \equiv  {Q_2 \over Q_1^2} -1,
  \label{eq-c1c2}
\end{equation}
where $ Q_1$, and  $ Q_2 $ are the
first- and second-order moments of $P_n$ for $N$ data points,
respectively, written as:
\begin{equation}
  Q_1 \equiv {1\over N}\sum_{n=n_0}^{n_0 + N-1} P_n,\ \ \ 
  Q_2 \equiv {1\over N}\sum_{n=n_0}^{n_0 + N-1} (P_n)^2.
\end{equation}
Here, note that $ Q_1$ is an averaged power for 
$ M \times N$ time-frequency components.
On the other hand, $ c_2 $ is defined by the second-order moment
normalized by the averaged power.
This value is analogous to the kurtosis (defined by 
the fourth-order moment of data),
which describes any non-Gaussianity of the data
\cite{bun-kur, bun-NRC}.

\subsection{Statistics of $ Q_1$ and $ Q_2$}

We calculate the statistics of parameters $ Q_1$ and $ Q_2 $
as a preparation for calculating the statistics of 
the parameters $ c_1$ and $ c_2 $, defined in Eq.\,(\ref{eq-c1c2}).
Although the $ M \times N$ components are not independent in practice,
because of overlapping in time and a window function,
we describe the following calculations while assuming that they are 
independent, for simplicity.
In practical use, the statistics (the averages, the variances, 
and the covariance) are estimated by 
replacing the time and frequency window size, $N$ and $M$, 
by effective time and frequency range, $N_{\rm eff}$ and $M_{\rm eff}$.
The effective window sizes, $N_{\rm eff}$ and $M_{\rm eff}$, are
estimated by simulations with Gaussian noises.

From Eq.\,(\ref{eq-P_mn_sum}), $ Q_1$ is also written 
by the sum of the noise, signal, and their correlation terms.
Thus, the expected value is
\begin{equation}
   E(Q_1)=
     {1\over N} \sum_n 
     \left\{{ S_n + E(N_n) +2 E(C_n) }\right\}
      = \alpha+1,
\end{equation}
where we define the average of the signal component power by
\begin{equation}
    \alpha \equiv {1\over N}\sum_n
    S_n,
\end{equation}
and we use relations $ E(N_n)=1$ and $ E(C_n)=0$.
The expected value of the square of $Q_1$ is written as
\begin{eqnarray}
   E(Q_1^2)&=&
     E\left({{1\over N^2}\sum_j  \sum_l 
     \left({ S_j + N_j +2 C_j }\right)
             \left({ S_l + N_l +2 C_l }\right) }\right)
        \nonumber \\
      &=& {2\alpha+1 \over M N} + (\alpha+1)^2,
         \nonumber
\end{eqnarray}
where we use relations $ E(N_k N_l) = E(N_k) E(N_l)$ $(k\neq l)$, 
$ E(N_n^2)=(M+1)/M$, and so on.
Thus, the variance of $Q_1$ is written as
\begin{equation}
   \mu_2(Q_1) = E(Q_1^2)-E(Q_1)^2
     =  {(2\alpha+1) \over M N}. 
\end{equation}

On the other hand, the expected value of $ Q_2$ is
\begin{eqnarray}
   E(Q_2) &=& E\left({{1\over N}\sum_n 
     \left({ S_n + N_n +2 C_n }\right)^2}\right)
        \nonumber \\
      &=& \beta_2\alpha^2+2\alpha+1 
            + {2 \alpha +1 \over M},
\end{eqnarray}
where $ \beta_2$ is a constant value related to the second-order 
moment of the signal,
\begin{equation}
   \beta_2 \alpha^2 = \sum_n S_n^2 /N . 
\end{equation}
Similarly, a constant value, $ \beta_3$,  is written as
$ \sum_n S_n^3 /N = \beta_3 \alpha^3$.
The constant numbers $\alpha$, $ \beta_2$ and $ \beta_3$
are determined only by the waveform and the amplitude of the signal.
The value $ \alpha $ represents the normalized signal power.
The value $ \beta_2$ depends on the time scale of the signal;
$ \beta_2$ becomes large for a short signal.

With more complicated, but similar, calculations, 
the variance of $Q_2$ and the covariance 
between $ Q_1$ and $ Q_2$ are obtained to be
\begin{eqnarray}
   \mu_2(Q_2)
     &=& {8 \over N} \beta_3\alpha^3
       + {20 M +32 \over M^2 N} \beta_2\alpha^2 \nonumber \\ 
      &&  + {16(M^2+M+3) \over M^3 N} \alpha \nonumber \\ 
      && + {2(2 M^2+5 M+3) \over M^3 N}, 
       \nonumber \\
   \mu_{11}(Q_1, Q_2)
     &=& {2 \over M N}
       \left\{ { \vphantom{\frac{(M)}{M}} 
        2 \beta_2\alpha^2  }\right. \nonumber \\
     &&  \left. { + {3(M + 1) \over M} \alpha
       + {M +1 \over M} }\right\}.
\end{eqnarray}

\subsection{Statistics for $c_1$ and $c_2$}

From the results described above, we obtain the expected value
and the variance of $ c_1 $ as
\begin{eqnarray}
   E(c_1) &=& \alpha +1 , \nonumber \\
   \mu_2(c_1) &=& {2\alpha+1 \over M N}.  
\end{eqnarray}

On the other hand, the expected value and the variance of 
$ c_2 $ are obtained as \cite{bun-kur}
\begin{eqnarray}
  E(c_2)&=&H_0 +O\left({1\over N}\right), \nonumber  \\
  \mu_2(c_2)&=&\mu_2(Q_1)H_1^2+ 2\mu_{11}(Q_1, Q_2) H_1 H_2 
      + \mu_2(Q_2) H_2^2  \nonumber  \\
    &&   +O\left({1\over N^{3/2}}\right),
    \label{eq-Dfun}
\end{eqnarray}
where
\begin{eqnarray*}
   H_0&=&c_2(E(Q_1),E(Q_2)),
     \\
   H_1&=& \left.{\partial c_2\over \partial Q_1
          }\right|_{Q_1=E(Q_1),Q_2=E(Q_2),}
       \\
   H_2&=& \left.{\partial c_2\over \partial Q_2
          }\right|_{Q_1=E(Q_1),Q_2=E(Q_2).}
\end{eqnarray*}
Thus, we obtain the mean and variance of $c_2$,
\begin{eqnarray}
   E(c_2) &=&{(\beta_2-1)\alpha^2 +{2\alpha +1 \over M}
    \over (\alpha+1)^2}, \ \\
   \mu_2(c_2)&=&{ 4\alpha^2 \over M N (\alpha+1)^6} \times \nonumber \\
    &&
   \left\{ {
      2 (\beta_3-\beta_2^2)\alpha^3  
     + (4\beta_3-3\beta_2^2-\beta_2) \alpha^2 }\right. \nonumber \\
     && \left.{  
       + 2(\beta_3-\beta_2) \alpha
       + (\beta_2-1)  }\right\},
\end{eqnarray}
and the covariance of $c_1$ and $c_2$,
\begin{eqnarray}
   \mu_{11}(c_1, c_2) ={ 2 (\beta_2-1) \alpha^2 \over M N (\alpha+1)^3}.
\end{eqnarray}
Here, we have neglected the higher terms, such as $O(1/M^2N)$, $O(1/M^3 N)$.

\subsection{Veto method}

\begin{figure}[t]
  \begin{center}
   \psfig{file=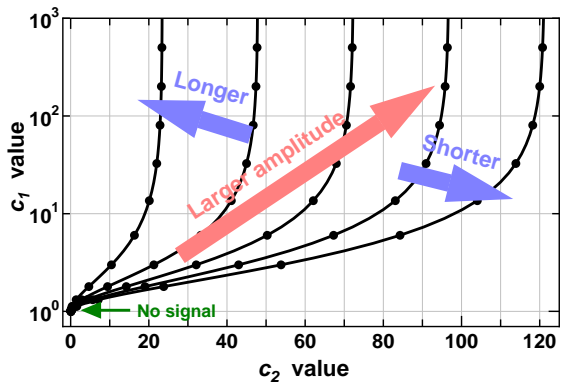,width=7.5cm}
   \caption{Theoretical predictions of the data point on 
     a $c_2$-$c_1$ plot for given waveform parameters 
     ($ \alpha$ and $\beta_2$).
      The loci corresponds to the $ \beta_2$ parameter of
      $\beta_2 = 122.21 \times  (0.2, 0.4, 0.6, 0.8, 1)$.
      The locus for a waveform with a large $\beta_2$ value 
      (with a short time scale) appears 
      on the right side of the plot.}
   \label{fig-t-model}
  \end{center}
\end{figure}

From the calculations described above, we can estimate
the expected values as $ E(c_1)$ and $ E(c_2)$
when the waveform and amplitude are given.
Figure\,\ref{fig-t-model} shows the expected points 
for given waveforms in the $c_2$-$c_1$ plane;
each curve is plotted by sweeping the power $\alpha$.
When $ \alpha $ is small, the ($c_2$, $c_1$) point is
around (0, 1) independently of the waveform ($\beta_2$).
(Here, we assume $M\gg1$.)
On the other hand, the position has a strong dependence 
on the $ \beta_2$ parameter for large $\alpha$:
$ c_2 \rightarrow \beta_2 -1$, for $\alpha \gg 1$.
Since signals with different time scales have different $\beta_2$ 
values, they appear along different loci.

Since the average time, $\Delta T$, is finite,
and since the data contains the Gaussian noise components, 
the ($c_2$, $c_1$) data 
points have a distribution around the predicted curve shown in
Fig.\,\ref{fig-t-model}. 
With an approximation that $\Delta T$ (or data point number $N$)
is sufficiently large, the ($c_2$, $c_1$) points for given
$\alpha $ and $\beta_2$ have a two-dimensional Gaussian distribution,
determined by the expected values ($ E(c_1)$ and $ E(c_2)$), 
the variances ($ \mu_2(c_1) $ and $ \mu_2(c_2)$), 
and the covariance $\mu_{11}(c_1, c_2)$.
We define the distance between a reference point, 
which is calculated by a gravitational waveform
(a reference waveform),
and a data point by
\begin{eqnarray}
   D^2&=&{1\over V} \left\{{ 
     \mu_2(c_2) \Delta c_1^2 }\right. \nonumber \\
    & & \left.{ -2 \mu_{11}(c_1,c_2) \Delta c_1 \Delta c_2
     + \mu_2(c_1) \Delta c_2^2
    }\right\},
   \label{eq-D2}
\end{eqnarray}
where $ \Delta c_1 $, $ \Delta c_2 $, and $ V $ are defined by
 $ \Delta c_1=c_1-E(c_1)$, $ \Delta c_2=c_2-E(c_2) $, 
and $ V= \mu_2(c_1)\mu_2(c_2)-\mu_{11}(c_1,c_2) $,
respectively (Fig.\,\ref{fig-model}).
This distance, which is normalized by the variances and covariance,
represents how similar these points are;
if $ D $ is small, the data point has a similar amplitude and
a time scale as that of the reference point.
Here, the minimum distance ($ D_{\rm min}$) for various 
amplitudes ($\alpha $) is the similarity of the time scales
of the data point and the reference waveform.
Thus, we use $D_{\rm min}$ to distinguish fakes
from the true GW signals.
If the data points have a two-dimensional Gaussian distribution
around their mean point, and if the variances and covariance are
sufficiently smaller than the curvature of the locus, 
the minimal distance ($D_{\rm min}$) approximately obeys 
an exponential distribution, 
\begin{equation}
  P(D_{\rm min}) \propto e^{-D_{\rm min}^2/2}.
\end{equation}

For a practical implementation of the veto scheme, we should
consider the acceptability of multiple reference waveforms
and a reduction of the computational load.
Thus, we adopted a conservative way for the veto analysis;
we reject only the events with longer time scales than the longest
one in the reference waveforms (Fig.\,\ref{fig-table}).
We use one waveforms with the smallest $\beta_2$ value,
i.e. with the longest time scale, as the reference waveform,
and set $D_{\rm min}=0$ if the data point is on the right side of 
the reference locus in the $c_1$-$c_2$ plot.
In addition, we prepare a map of $ D_{\rm min}$ in the $c_1$-$c_2$ 
plane before the data analysis, so as to reduce the 
computational load during the data analysis.


\end{document}